\documentclass[aps,pra,floatfix,showpacs,preprint,superscriptaddress]{revtex4} 

\usepackage{amsmath} 
\usepackage{graphicx} 
\usepackage{epsfig}

\newcommand{\ket}[1]{\mbox{$|#1\rangle$}}

\newcommand{\op}[1]{\mbox{\boldmath $\hat{#1}$}}

\begin{document} 

\title{Conditional phase shifts using trapped atoms} 

\author{A.~Gilchrist and G.~J.~Milburn$^*$ } 
\affiliation{Centre for Quantum Computer Technology and Department of Physics, The University of Queensland, 
St Lucia, QLD, 4072 Australia.\\
 $\mbox{}^*$~Department of Applied Mathematics and Theoretical Physics, University of Cambridge, Cambridge,UK }

\date{\today}

\begin{abstract} 
  We describe a scheme for producing conditional nonlinear phase shifts on
  two-photon optical fields using an interaction with one or more ancilla
  two-level atomic systems.  The conditional field state transformations
  are induced by using high efficiency fluorescence shelving measurements
  on the atomic ancilla. The scheme can be nearly deterministic and is of
  obvious benefit for quantum information applications.
\end{abstract}


\maketitle

It has recently been shown \cite{KLM} that nonlinear phase shifts on two
photon states can be produced by coupling the mode of interest to ancilla
modes via a beam splitter and making photon counting measurements on the
ancilla modes.  Such conditional nonlinear phase shifts can be used to
perform two qubit operations for logical states encoded in photon number
states. If such conditional gates are used to prepare entangled states for
teleportation, efficient quantum computation can be performed which with
suitable error correcting codes can be made fault tolerant \cite{KLM}.  In
this paper we show that if the ancilla modes are replaced with a two level
atom similar conditional nonlinear phase shifts can be achieved by near
deterministic postslection on atomic measurements.  The atomic measurements
can made with fluorescence shelving techniques which are very much more
efficient than single photon counting measurements, thus reducing the need
for new photon counting technologies inherent in the KLM scheme. 
Recently, a different scheme for conditional quantum gates based on atomic systems
was presneted by Protsenko et al.\cite{protsenko}

Consider a single optical mode prepared in an arbitrary two photon state
\begin{equation}
|\psi\rangle=c_0|0\rangle+c_1|1\rangle+c_2|2\rangle
\end{equation}
Our objective is to find a way to produce the nonlinear phase shift
transformation defined by
\begin{equation}
|\psi\rangle\rightarrow c_0|0\rangle+c_1|1\rangle-c_2|2\rangle
\end{equation}
To achieve this result we assume that at some fixed time an interaction
between the field mode and a single two level atom is switched on. After
some interaction time $t$ the interaction is turned off and the atomic
state is measured by fluorescence shelving. The resulting conditional state
of the field will then depend on the initial state of the two-level atom
and the interaction time. We will show that these can be so arranged as to
effect the nonlinear phase shift required.

We have in mind a quantum computing communication protocol in which the
optical field mode is derived from a transform limited pulsed field which
is rapidly switched into the cavity mode containing the atomic systems at
fixed times determined by the pulse repetition rate.  Similar systems have
been proposed as a quantum memory for optical information processing
\cite{Pittman2002}.  When the atomic measurement yields the required result
the field may be switched out again for further analysis or subsequent
processing through linear and conditional elements.

\begin{figure}
\centerline{\epsfig{file=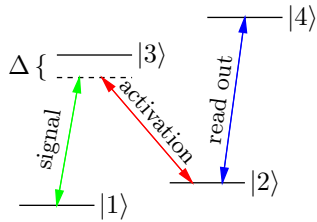}}
\caption{Level scheme for an effective two-level transition controlled by a
stimulated Raman process.}
\label{fig1}
\end{figure}

Once the cavity field is prepared, we need to switch on the interaction
with the atomic system. In order that we can switch this interaction at
predetermined times we propose that an effective two level transition
connected by a Raman process with one classical field and the quantised
signal field, be used.  A similar scheme has recently been proposed as the
basis of a high efficiency photon counting measurement
\cite{Imamoglu,James_Kwiat}. The process is also used in the EIT schemes
for storing photonic information \cite{Lukin2000} and for quantum state
transfer between distant cavities \cite{Cirac1997}. The level diagram is
shown in figure \ref{fig1}. The nearly degenerate levels $|1\rangle$ and
$|2\rangle$ are connected by a stimulated Raman transition to level
$|3\rangle$. The detuning of the Raman pulse from the excited state
$|3\rangle$ is $\Delta$, which is approximately the same as the detuning of
the signal mode form the same transition.  An advantage of using a
stimulated Raman process of this kind is that the excited state $|2\rangle$
can be a metastable, long lived level. We thus do not need to consider
spontaneous emission from this level back to the ground state.  The readout
of the atomic system may be achieved by using a cycling transition between
the excited state $|2\rangle$ and another probe level $|4\rangle$.  Such
measurements are routinely performed in ion trap studies \cite{Rowe2000}
and can have efficiencies greater than 99\%.

The interaction between the single mode field and a two-level atom is
described by the effective Hamiltonian
\begin{equation} 
\label{eq:1} 
\op{H}=\kappa(\op{a}^\dagger \op{\sigma}^-+\op{a}\op{\sigma}^+) 
\end{equation} 
The interaction strength is given by $\kappa=\Omega g/2\Delta$ where
$\Omega$ is the Rabi frequency for the Raman pulse and $g$ is the one
photon Rabi frequency for the signal field.  The unitary transformation
that acts when this interaction is applied for a time $t$ is
\begin{equation} 
\label{eq:1.1} 
\op{U}(\tau)=\exp[-i\tau(\op{a}^\dagger \op{\sigma}^-+\op{a}\op{\sigma}^+)] 
\end{equation} 
where $\tau=\kappa t$.

If the atom is prepared in the ground state and found in the ground state 
 after the interaction, the conditional state of the field is given by 
\begin{equation} 
\label{eq:2} 
\op{\Upsilon}_{gg}(\tau)\ket{\psi} = 
\cos(\tau\sqrt{\op{a}^\dagger\op{a}})\ket{\psi} 
\end{equation} 
On the other hand if the atom is prepared in the exited state and found in 
the excited state after the interaction, the conditional state is given by 
\begin{equation} 
\label{eq:3} 
\op{\Upsilon}_{ee}(\tau)\ket{\psi} = 
\cos(\tau\sqrt{\op{a}\op{a}^\dagger})\ket{\psi} 
\end{equation} 
There is considerable practical advantage to using the excited state
preparation rather than the ground state as there is always a signal for
correct operation, however the analysis is the same.

Now suppose we send in a generic two photon state 
\begin{equation} 
\label{eq:4} 
\ket{\psi}=c_0\ket{0}+c_1\ket{1}+c_2\ket{2} 
\end{equation} 
which interacts with a single two level atom, prepared in the ground state, 
and after the interaction the atom is found still to be in the ground 
state. In this case we need to apply the measurement operator 
$\op{\Upsilon}_{gg}$, and the resulting state of the field is 
\begin{equation} 
\label{eq:5} 
\ket{\phi_g}=A_0c_0\ket{0}+A_1c_1\ket{1}+A_2c_2\ket{2} 
\end{equation} 
where $A_0=1$, $A_1=\cos(\tau)$ and $A_2=\cos(\sqrt{2}\tau)$. 
Note that the frequencies of the $A_1$ and $A_2$ terms are irrational 
multiples of each other. As we sweep $\tau$, $A_1$ and $A_2$ should 
explore their entire phase space. It should be possible to find values of 
$\tau$ for which $(A_1,A_2)$ approaches arbitrarily close to $(1,-1)$. 
These solutions can be found trivially for small $\tau$ by plotting 
$A_1$ and $-A_2$ and visually inspecting for intersections near 1. 
Some high-probability solutions are summarised in table~\ref{tab:Ygg}. 

\begin{table}[htbp] 
\begin{tabular}[c]{|c|ccc|}
\hline 
$\tau$ & $A_0$ & $A_1$ & $A_2$ \\ 
\hline 
6.5064 & 1 & 0.97519 & -0.97516 \\ 
37.73742 & 1 & 0.9992663 & -0.9992665 \\ 
219.918 & 1 & 0.999979 & -0.999978 \\ 
\hline 

\end{tabular}\centering 
\caption{High-probability results for a single atom initially in 
a ground state and measured in a ground state after an interaction time 
$\tau$.} 
\label{tab:Ygg} 
\end{table} 

With two atoms, one initially prepared in the ground state and the second 
prepared in the excited state, the conditional state given that both atoms are 
found in their initial state after the interaction is 
\begin{eqnarray} 
\label{eq:6} 
\ket{\phi_{ge}}&=&\op{\Upsilon}_{gg}(\tau_1) 
\op{\Upsilon}_{ee}(\tau_2)\ket{\psi}\\ 
&=&A_0c_0\ket{0}+A_1c_1\ket{1}+A_2c_2\ket{2} 
\end{eqnarray} 
where 
$A_0=\cos(\tau_2)$, $A_1=\cos(\tau_1)\cos(\sqrt{2}\tau_2)$ 
, and $A_2=\cos(\sqrt{2}\tau_1)\cos(\sqrt{3}\tau_2)$. 

By using two atoms, each with a different interaction time, we have more 
freedom in locating solutions for which the magnitude of the $A_i$'s are 
closer together. This occurs at the expense of a more complex experimental
 scheme. To find solutions we employed a 
simulated annealing algorithm on an initial ensemble of randomly chosen 
points. After the points where suitably `cooled' relative to a penalty 
function, we applied a simplex minimisation method on the top contenders to 
find the local minima. Some interaction times that result in implementing 
nearly ideal nonlinear-sign gates with high probability are given in 
table~\ref{tab:YggYee}. 

\begin{table}[htbp] 
\begin{tabular}[c]{|c|ccc|}
\hline 
$\tau_1,\tau_2$ & $A_0$ & $A_1$ & $A_2$ \\ 
\hline 
477.60911391, 197.78326606 & -0.9906204535 & -0.9906204532 & +0.9906204537 \\ 
37.79300921, 197.78109842 & -0.9903219354 & -0.9903219357 & +0.9903219350 \\
\hline 
\end{tabular}\centering 
\caption{High-probability results for two atoms, one initially in 
a ground state the other in an exited state and detected in their 
initial states after interaction times $\tau_1$ and $\tau_2$.} 
\label{tab:YggYee} 
\end{table} 

\begin{figure}
\centerline{\epsfig{file=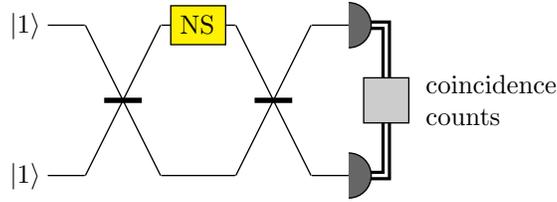}}
\caption{An interferometric scheme for calibrating the conditional nonlinear sign shift gate (NS) 
  by searching for coincidence counts from the photon detectors (PD).}
\label{fig2}
\end{figure}

In an experiment it would be necessary to find a way to calibrate the
interaction time until the desired phase shift had been reached. One way to
do this is depicted in figure \ref{fig2}.  Two single modes, each prepared
in a one photon state are incident on a 50/50 beam splitter.  The two
photon interference then results in the state
$|2\rangle|0\rangle-|0\rangle|2\rangle$.  The conditional phase shift can
then be inserted on one arm so that the state is transformed to
$|2\rangle|0\rangle-e^{i\theta}|0\rangle|2\rangle$. This state is then run
through an identical 50/50 beam splitter to the first and the probability
for coincidence counts is sampled. This probability is given by
$P_c=\cos^2\theta/2$. The coincidence detection rate then drops to zero at
the required conditional phase shift of $\theta=\pi$.

We now estimate some typical values for the parameters. In a recent
experiment a similar stimulated Raman process was observed using single
rubidium atoms falling through a high finesse optical cavity
\cite{Henrich2000}. The following parameters are typical of that
experiment: $g=2\pi\times 4.5$~MHz, $\Omega=2\pi\times 30 $~MHz and
$\Delta=2\pi\times 6$~MHz. This gives a coupling constant of the order of
$70$~Mhz.  To achieve effective interaction constants of the order of those
in the table \ref{tab:Ygg} requires interaction times of the order of
$0.1-5$~$\mu$s.  In this paper we have neglected cavity decay which
obviously needs to be kept small over similar time scale, which while
difficult is not impossible.

We would like to thank the Computer Science department of the University of 
Waikato for making available their computing resources. 
AG was supported by the New Zealand Foundation for Research, Science 
and Technology under grant UQSL0001. GJM was supported by the Cambridge-MIT Institute
while a visitor at University of Cambridge. 
\bibliographystyle{prsty}

\end{document}